\documentclass[twocolumn,linen]{aastex631}
\usepackage{{booktabs}}
\usepackage{romannum}
\usepackage{float}
\usepackage{hyperref}
\usepackage{soul}
\usepackage{xfrac}
\setulcolor{red}
\setstcolor{red}

\def\rsun{R$_{\odot}$}

\newcommand{\ca}{\mbox{Ca\,{\sc ii}~K\,}}
\newcommand{\ccmax}{CC$_{\rm max}$}

\begin{document}

\title{Differential Rotation of the Solar Chromosphere: A Century-long Perspective from Kodaikanal Solar Observatory \ca\ Data}
\shorttitle{Differential Rotation of the Solar Chromosphere}
\shortauthors{D. K. Mishra et al.}

\correspondingauthor{Dipankar Banerjee}
\email{dipu@aries.res.in}

\author[0009-0003-1377-0653]{Dibya Kirti Mishra}
\affiliation{Aryabhatta Research Institute of Observational Sciences, Nainital-263002, Uttarakhand, India}
\affiliation{Mahatma Jyotiba Phule Rohilkhand University, Bareilly-243006, Uttar Pradesh, India }

\author[0009-0008-5834-4590]{Srinjana Routh}
\affiliation{Aryabhatta Research Institute of Observational Sciences, Nainital-263002, Uttarakhand, India}
\affiliation{Mahatma Jyotiba Phule Rohilkhand University, Bareilly-243006, Uttar Pradesh, India }

\author[0000-0003-3191-4625]{Bibhuti Kumar Jha}
\affiliation{Southwest Research Institute, Boulder, CO 80302, USA}

\author[0000-0002-0335-9831]{Theodosios Chatzistergos}
\affiliation{Max Planck Institute for Solar System Research, Justus-von-Liebig-Weg 3, D-37077 Göttingen,Germany}

\author[0000-0001-7570-545X]{Judhajeet Basu}
\affiliation{Indian Institute of Astrophysics, Koramangala, Bangalore 560034, India}

\author[0000-0002-5014-7022]{Subhamoy Chatterjee}
\affiliation{Southwest Research Institute, Boulder, CO 80302, USA}

\author[0000-0003-4653-6823]{Dipankar Banerjee}
\affiliation{Aryabhatta Research Institute of Observational Sciences, Nainital-263002, Uttarakhand, India}
\affiliation{Indian Institute of Astrophysics, Koramangala, Bangalore 560034, India}
\affiliation{Center of Excellence in Space Sciences India, IISER Kolkata, Mohanpur 741246, West Bengal, India}

\author[0000-0003-2596-9523]{Ilaria Ermolli}
\affiliation{INAF Osservatorio Astronomico di Roma, Via Frascati 33, 00078 Monte Porzio Catone, Italy}

\begin{abstract}

Chromospheric differential rotation is a key component in comprehending the atmospheric coupling between the chromosphere and the photosphere at different phases of the solar cycle. In this study, we therefore utilize the newly calibrated multidecadal \ca\ spectroheliograms (1907--2007) from the Kodaikanal Solar Observatory (KoSO) to investigate the differential rotation of the solar chromosphere using the technique of image cross-correlation. Our analysis yields the chromospheric differential rotation rate $\Omega (\theta) = (14.61\pm 0.04 - 2.18\pm 0.37\sin^2{\theta} - 1.10 \pm 0.61\sin^4{\theta})^\circ{\rm /day}$. These results suggest the chromospheric plages exhibit an equatorial rotation rate 1.59\% faster than the photosphere when compared with the differential rotation rate measured using sunspots and also a smaller latitudinal gradient compared to the same. To compare our results to those from other observatories, we have applied our method on a small sample of \ca\ data from Rome, Meudon, and Mt. Wilson observatories, which support our findings from KoSO data.
Additionally, we have not found any significant north-south asymmetry or any systematic variation in chromospheric differential rotation over the last century.
\end{abstract}


\section{Introduction} \label{sec:intro}

Solar rotation has been one of the persistent topics of interest in solar physics since its discovery at the beginning of the 17th century \citep{paterno_solar_2010}. Early investigations into this phenomenon were primarily based on the tracking of prominent dark photospheric magnetic features called sunspots \citep{Carrington1859,Newton1951}, which enabled us to measure the photospheric differential rotation giving rise to the well-known empirical relation
\begin{equation}
\Omega= A + B\sin^2{\theta} +C \sin^4{\theta},
\label{eq1}
\end{equation}
where $\theta$ is the latitude, $A$ is the equatorial rotation rate while $B$ and $C$ are coefficients  of a quadratic expansion in $\sin^2{\theta}$. In the last century, there has been an outstanding advancement in instruments and measuring techniques, which not only improved the method of sunspot tracking \citep{Newton1951,Ward1966,Balthasar1986,gupta1999,javaraiah2005,Jha2021} but also led us to measure the solar rotation based on new techniques such as spectroscopy \citep{Howard1970, Howard1984} and helioseismology \citep{Komm2008, Howe2009}. As a result, in the last few decades, extensive research has been conducted in this field, leading to a consensus among researchers about the differential rotation profile of the Sun in both its interior and photosphere. Despite such extensive work, many questions still need to be answered. One of many such prevailing questions is the variation in the rotational profile of the higher solar atmosphere, where the magnetic field mainly dominates the dynamics \citep{Stix1976,Gary2001,Gomez2019}.

Observations on spectral lines like \ca\ centered at $393.367$\,nm, probe the chromospheric layer of the Sun \citep{linsky_solar_1970,Livingston2007}, thus opening up new doors directed towards answering the question of chromospheric differential rotation. One of the predominantly visible features in the \ca\ observations is the chromospheric plage, which is generally found above sunspots. These plages are the large-scale magnetic structures \citep{zirin_1974} having a relatively extended lifetime compared to the other features observed in \ca\ observations. The extended lifetime and relatively stable nature of these plages make them an ideal candidate for measuring the chromospheric rotational profile \citep{Singh1985, Bertello2020}.

Chromospheric plages are, however, extended structures \citep[extend up to $\approx200,000$\,km;][]{priest2014} compared to sunspots \citep[up to $\approx 60,000$\,km;][]{solanki2003} that seem to change their morphology relatively faster when compared to the same. This makes tracer-based tracking algorithms, similar to the one used in, e.g. \citet{Newton1951}, \citet{Jha2021} etc., unreliable for application. Having recognized this problem, \citet{Livingston1969,Livingston1979} utilized H$\alpha$ spectroscopic data for the chromosphere and reported a rotation rate $3\%-8$\% faster than the underlying photosphere. Despite the limitations of the tracer method, some have attempted to use it to investigate chromospheric differential rotation. \citet{schroter1975,belvedre1977,antonucci1979,Ternullo1987,brajsa1991} employed this approach by tracing the \mbox{Ca\,{\sc ii}} network, plages, and polar filaments. The collective findings from these studies indicate that the chromosphere rotates $1\%-5$\% faster than the photosphere. \citet{ternullo_rotation_1986} also reported variations in the rotation of plage with their age. In addition to these, recent studies conducted by \citet{Li2020,xu2020}, utilizing data in \mbox{He\,{\sc i}} and  \mbox{Mg\,{\sc ii}} lines, reported a faster rotation of the chromosphere. Subsequently, many have extended their observations to higher layers of the solar atmosphere \citep{Brajsa2004,Chandra2010,Li2019,Sharma2020,Zhang2023}, and employed alternative methods, including the tracing of Coronal Bright Points \citep[CBPs;][]{Brajsa2004,hara2009differential,simon1972solar}, as well as the Lomb-Scargle periodogram analysis \citep{Li2019} and autocorrelation method \citep{Chandra2010,Sharma2020,Zhang2023}. The outcomes from these studies also suggest the faster rotation of the higher solar atmosphere compared to the photosphere.

These results were contrasting those by \citet{Singh1985}, who employed the Fast Fourier Transform (FFT) method on \ca\ plage area data (1951--1981) obtained at Kodaikanal Solar Observatory (KoSO) and concluded that the chromosphere exhibited a slower rotation compared to the photosphere. In a recent study, a different approach was employed to support this observation further. \citet{Bertello2020} employed the image cross-correlation technique, instead of FFT, on \ca\ data acquired at Mount Wilson Observatory (MWO; 1915--1985) and found that the chromospheric plages give 0.63\% slower rotation than the photospheric sunspots. Furthermore, studies comparing active regions in the photosphere and their counterparts in the chromosphere highlight their great spatial correspondence \citep[e.g.][]{babcock_suns_1955,loukitcheva_relationship_2009,chatzistergos_recovering_2019,murabito_investigating_2023}, which would suggest that there is no significant change in the rotational profile between photosphere and chromosphere.

The conflicting findings among these various studies pose a significant challenge in establishing a definitive understanding of the rotation profile of the chromosphere in relation to the underlying photosphere. Consequently, the question remains: Does the chromosphere rotate faster, slower, or in the same way as the photosphere? To gain deeper insights into the chromospheric rotation profile and its connection with the underlying photosphere, it is crucial to have a comprehensive and consistent dataset that minimizes biases caused by different aspects of solar activity, such as the phase and strength of the solar cycle. Fortunately, the Kodaikanal Solar Observatory~(KoSO) possesses an extensive collection of \ca\ archival data spanning over a century (1904--2007), obtained using the same setup throughout this period \citep{priyal2014, Jha2022thesis}. These data have recently been re-calibrated by \citet{theo2018, chatzistergos_analysis_2020}, resulting in improved data quality. The availability of such long-term data presents a significant advantage, as it allows differential rotation measurement and temporal variation over the last century and helps us to resolve some of the questions about chromospheric rotation. This article presents an overview of the data and their processing in \autoref{Sec:data}. 
In \autoref{Sec:method}, we describe our approach to computing the differential rotation and present our results in \autoref{Sec:results}. Finally, we discuss our results and summarize our conclusions in \autoref{Sec:Discussion} and \autoref{Sec:summary}.

\section{Data and processing} \label{Sec:data}

The KoSO has a rich collection of \ca\ spectroheliograms taken on photographic plates/films, spanning over a century from 1904 to 2007 and is one of the oldest repositories of such data \citep{chatzistergos_full-disc_2022,Jha2022thesis}. At KoSO, a spectroheliograph with a 30\,cm objective lens and f/21 focal ratio was used for these observations. This spectroheliograph is fed by a siderostat, which compensates for the effect of the rotation of the Earth by keeping the reflected beam of sunlight in a fixed direction. Later, this reflected beam is passed through a diffraction grating system that allows \ca\ wavelength with a pass band of 0.05\,nm centered at 393.367\,nm  \citep{Bappu1967, Jha2022thesis}. In recent years, these photographic plates/films have been digitized using a 4096\,$\times$\,4096 pixels CCD sensor, with a bit depth of 16-bit and made available to the wider scientific communities\footnote{The digitized data can be accessed through \url{https://kso.iiap.res.in/data.}} \citep{priyal2014, Chatterjee2016}.

There have been various studies analyzing the KoSO \ca~data \citep[e.g.][]{priyal2014,Chatterjee2016}.
However, a more accurate calibration method was recently developed by \citet{theo2018} and applied to the KoSO Ca II K observations, resulting in an improved series of these data. A brief summary of the method implemented in \citet{theo2018} is as follows. This calibration involves several steps, which are pre-processing, photometric calibration and limb-darkening compensation. The main part of the pre-processing is the detection and circularisation of the solar disk in the images. Then, the photometric calibration is performed by constructing the calibration curve for each image accomplished by comparing the observed quiet Sun center-to-limb variation with that from modern CCD-based observations. Finally, a limb-darkening correction is applied to obtain the contrast images with uniform intensity over the solar disc up to 0.99 of its radius. See \citet{theo2018,theo2019,theo2019SoPh,chatzistergos_analysis_2020} for the comprehensive details of all these processing steps. In addition to these calibration processes, a novel and precise method was developed by \citet{Jha2022thesis} to orient KoSO images accurately. 
We note that accurate knowledge of the observation time is needed to align the images correctly.
However, the data from the KoSO was found to have the inconsistent time of observations, resulting in erroneous orientation. \citet{Jha2022thesis} successfully addressed these issues by resolving the inconsistencies in the time of observation of the images. Here, for the purposes of the current work, we utilized the recently calibrated \citep{chatzistergos_analysis_2020} and correctly oriented observations \citep{Jha2022thesis} of \ca\ for the period of 1907--2007. We emphasise that the data obtained during the period of 1904--1906 are not used in this work due to the unavailability of the timestamp information for the mentioned period, which is required to correctly orient the images using the method given in \citet{Jha2022thesis}. Representative examples of calibrated and orientation-corrected images are shown in \autoref{fulldisk}(a) and \ref{fulldisk}(b).

Besides the KoSO \ca\ data, we have also made use of a small sample of other \ca\ observations from a few relevant archives for comparison purposes.
In particular, we used data from the Meudon \citep{malherbe2019}, Mt Wilson \citep{Bertello2020}, and Rome \citep[taken with the Rome Precision Solar Photometric Telescope, Rome/PSPT;][]{ermolli10.3389} datasets.
Meudon is one of the oldest \ca\ archives, with observations since 1893 and continues to this day after some modifications in the instrumentation  \citep{malherbe_130_2023}. These were photographic observations stored on glass plates up to 27 September 2002, while observations with a CCD camera started on 13 May 2002, and they were performed with a spectroheliograph with a nominal pass band of 0.015nm \citep{malherbe2019}.
Mt Wilson also comprises spectroheliograms covering the period 1915--1985 with a nominal pass band of 0.035\,nm \citep[][]{tlatov2009}.
Rome/PSPT has observations from 1996 up to the present. It uses a CCD camera and an interference filter with a bandwidth of 0.25nm \citep[][]{ermolli10.3389}.
In particular, we used 829 and 569 images from Meudon and Rome/PSPT, respectively, over the period of 2000\,--\,2002, which is close to the solar maximum of cycle 23 and 408 images from Mt Wilson over the period 1978--1979, in the ascending phase of cycle 21.
The images were processed with the same methods as KoSO to compensate for the limb darkening and perform the photometric calibration for the photographic data from Meudon and Mt Wilson \citep{theo2018,theo2019,chatzistergos_analysis_2020}.

Finally, for comparison purposes, we also used Mt Wilson data over the period 1978--1979 processed by \citet{bertello2010SoPh}\footnote{Available at \url{ftp://howard.astro.ucla.edu/pub/obs/CaK/run_mean_flat/}}, as well as raw (without processing to compensate the limb-darkening or perform the photometric calibration) Mt Wilson data.
For the raw data, we further used two different versions; the first one is with the preprocessing (definition of center coordinates and image rotation) by \citet{bertello2010SoPh} and the second one by \citet{chatzistergos_analysis_2020}.
The latter case applied a correction for the recorded solar disk ellipticity, thus accounting for the image distortions \citep[see][]{chatzistergos_historical_2020,chatzistergos_analysis_2020}.

\section{Methodology}\label{Sec:method}

Considering the relatively dynamic and spatially extended nature of the plage region in \ca\ observations, in this work, we used the image cross-correlation-based technique to quantify the chromospheric differential rotation rate. We start with selecting a pair of observations, preferably consecutive, but in such a way that the difference in observation time ($\Delta t$) of the selected observation is more than 0.5\,days but less than 1.5\,days. This lower limit on $\Delta t$ is imposed to eliminate the effects of other rapidly evolving features (having a life span less than 0.5\,days) and the upper limit to minimize the effects of the evolution of plages. Furthermore, these limits have also been used to ensure that one image per day will be selected, which is close to the average cadence of the KoSO digital archive.

\begin{figure}[htb!]
\centering
\includegraphics[width=\columnwidth]{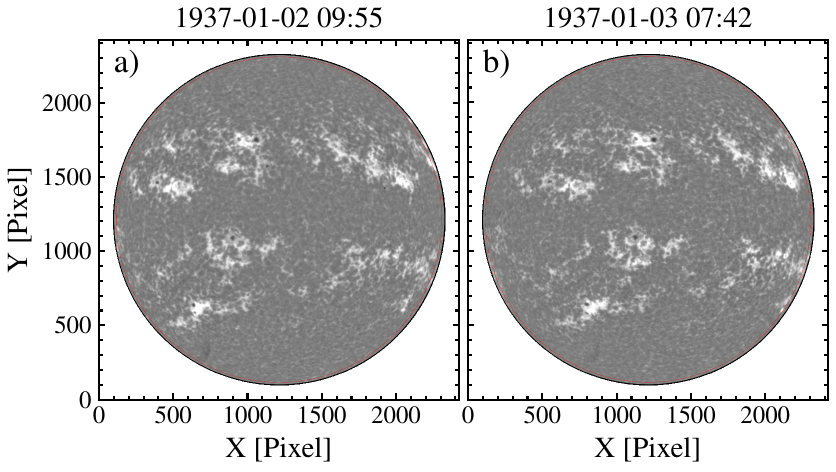}
\includegraphics[width=\columnwidth]{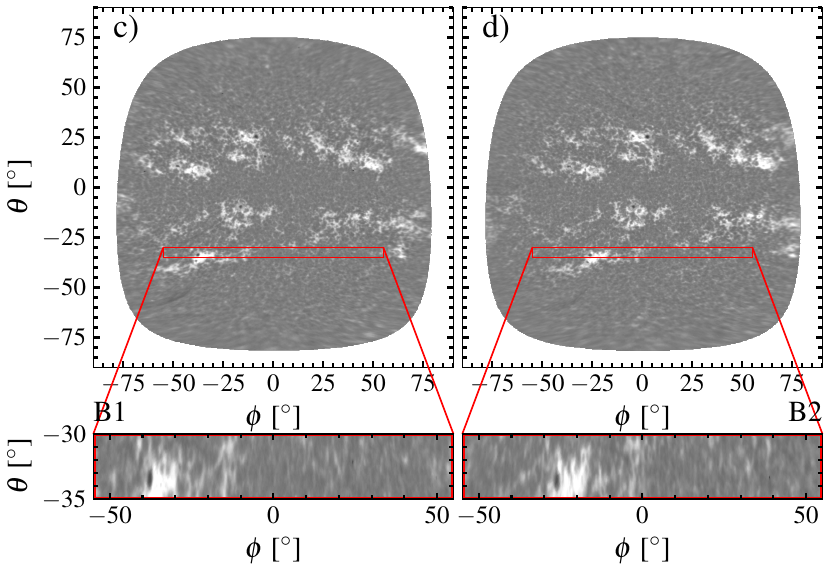}
\caption{Example pair of calibrated and rotation-corrected full disk \ca\ images from KoSO as recorded on (a) 1937-01-02 09:55 IST and (b) 1937-01-03 07:42 IST. The red dotted circles represent the 0.98\,\rsun limit. Panels (c) and (d) show the corresponding full disk images projected on the heliographic grid. Red rectangular boxes in (c) and (d) represent the selected latitude bands (in this case $-35^{\circ}$ to $-30^{\circ}$) for the cross-correlation. Zoomed-in views of the bands are shown in the inset (B1 and B2).}
 \label{fulldisk}
\end{figure}

Firstly, we project the selected pair of full disk observations, shown in \autoref{fulldisk}(a) and \ref{fulldisk}(b), to the heliographic grid \citep{Thompson2006} of size 1800\,pixels\,$\times$\,1800\,pixels ($0.1^\circ$/pixel in latitude and longitude) as shown in \autoref{fulldisk}(c) and (d), using the near-point interpolation method. While projecting the full disk observation on the heliographic grid, we restricted ourselves to the inner 0.98\,\rsun\ (shown by the red line in \autoref{fulldisk}(a) and (b)) to avoid the significant distortion near the limbs caused by the projection effects. 
Here, we have not reduced the dimension of the image but only analysed the disk up to 0.98\,\rsun. 
Then, we split the image in 5$^{\circ}$ latitudinal bands (see red rectangular boxes and zoomed-in view shown in insets in \autoref{fulldisk} B1 and B2). 
We restrict ourselves between $\pm55^{\circ}$ in latitude and longitude to further minimize the effect of projection near the limb and also because plages are very unlikely to appear above these latitudes. Following this, we apply the 2D image cross-correlation technique within the 5$^{\circ}$ latitudinal bands to get the magnitude of the spatial movement of the features (here, plages). The steps of image cross-correlations are described as follows.

\graphicspath{ {./figures/} }

\begin{figure}[htb!]
\centering
\includegraphics[width=\columnwidth]{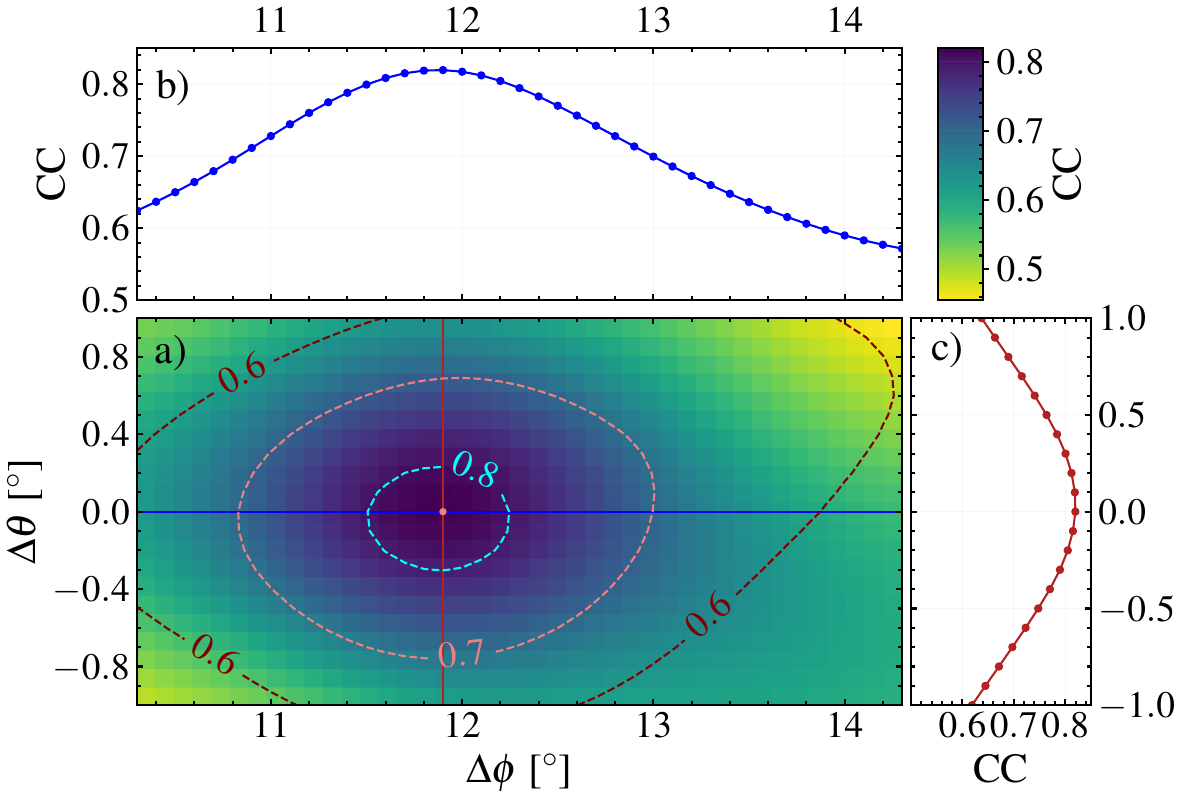}
\caption{ (a) Variation of the correlation coefficient for all the combinations of longitudinal ($\Delta\phi$) and latitudinal ($\Delta\theta$) shifts. The red dot represents the point of maximum CC (\ccmax), and the contours show the lines of constant CC. (b) and (c) show the variation of CC with $\Delta\phi$ for constant $\Delta\theta$ (along blue horizontal line) and,  $\Delta\theta$ for constant $\Delta\phi$ (along red vertical line), for $\Delta\phi$ and $\Delta\theta$ corresponding to the location of maximum CC.}
 \label{correlation1}
\end{figure}

To reduce the computational time, we start with the initial guess [$\Delta\phi_0$ and $\Delta\theta_0$] that presumably gives the best correlation coefficient. The $\Delta\phi_0$ is calculated based on the photospheric rotation rate measured from the KoSO White Light (WL) data \citep{Jha2021} for the selected band using $\Delta\phi_0 = \Omega(\theta)\Delta t$, where $\Omega(\theta)$ is the angular rotation rate at latitude $\theta$ (taken as the mid-latitude of the selected band). Assuming that the plages do not show considerable movement in the meridional plane, $\Delta\theta_0$ is taken as 0. We then calculate the standard 2D cross-correlation (CC)\footnote{The image cross-correlation was performed using correl\_images.pro routine available in the Solar SoftWare library. For detail see \url{https://hesperia.gsfc.nasa.gov/ssw/gen/idl_libs/astron/image/correl_images.pro.}} for each $0.1^\circ$ shift in the range of $[\Delta\phi_0\pm 2^\circ$, $\Delta\theta_0\pm 1^\circ$]. In \autoref{correlation1}, we plot the correlation matrix for all the combinations of shifts for B1 and B2 in physical units ($\Delta\theta$ and $\Delta\phi$, i.e., shift in latitude and longitude) by adjusting it for the initial offset of $[\Delta\phi_0$, $\Delta\theta_0$], for the better interpretation. Consequently, we calculate the optimum shift ($\Delta\theta$ and $\Delta\phi$) for which the CC value is maximum (represented by the small red dot in \autoref{correlation1}a). In \autoref{correlation1}(b) and (c), we also show the variation of CC with $\Delta\phi$ for fixed $\Delta\theta$ and with $\Delta\theta$ for fixed $\Delta\phi$, respectively. We repeat this process for all latitude bands in the range of $\pm55^\circ$ latitude and for all the pairs of observations from the period 1907--2007.

\begin{figure}[htb!]
\centering
\includegraphics[width=\columnwidth]{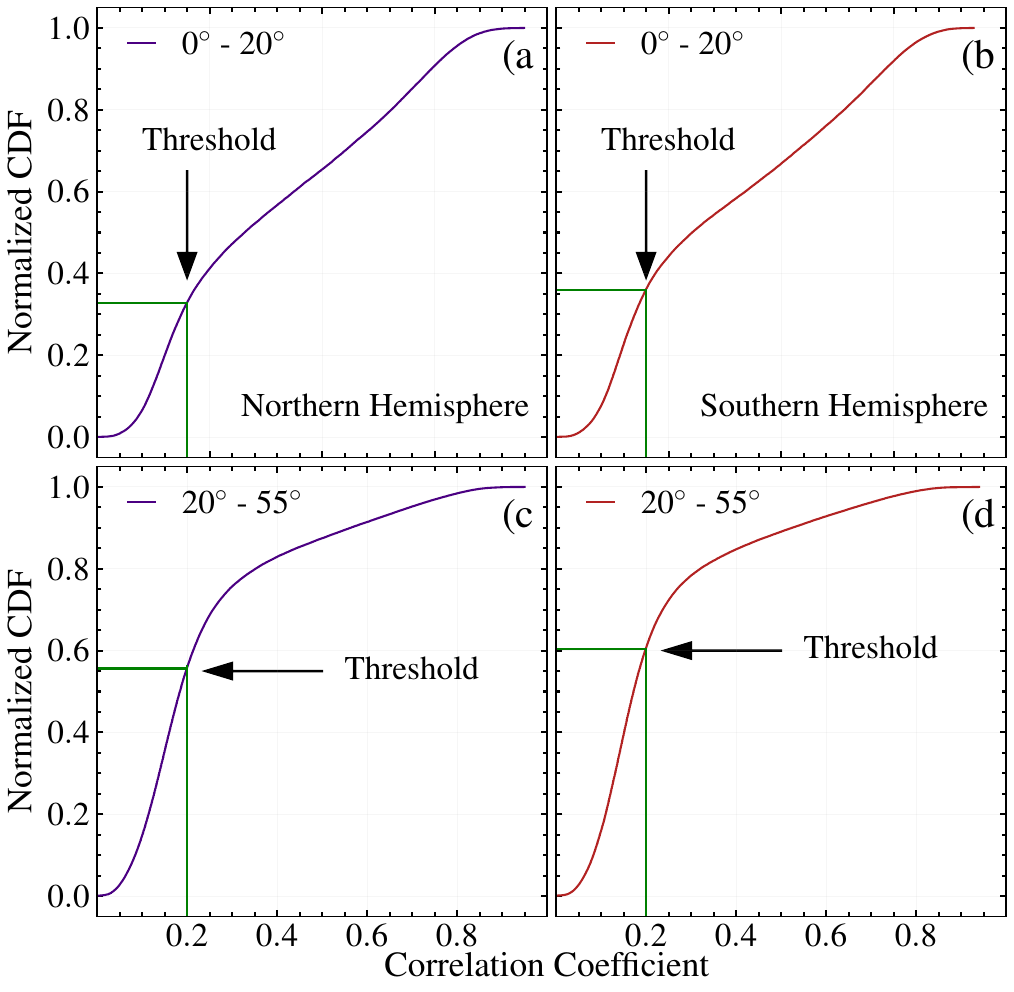}
\caption{The normalized cumulative distribution of \ccmax\ for the Northern hemisphere for latitude range (a) $\theta \leq 20^\circ$ and (b) $\theta > 20^\circ$. A similar plot for the Southern hemisphere is shown in  (b) and (d). The arrow in each panel highlights the lower threshold limit of \ccmax\, which is 0.2, used in our analysis.}
 \label{cor_coeff}
\end{figure}

We find for $\approx$\,50\% of cases ($\approx$\,66\% and $\approx$\,42\% for $\theta<20$ and $\theta>20$ respectively), considering all the latitude bands, the maximum correlation coefficient (\ccmax) lies in the range [0.2, 0.8]. Additionally, We have also encountered cases where (i) \ccmax\ is less than 0.2 and (ii) no local maximum (\ccmax) is found, i.e., either it lies at the extreme ends or no/minute variation in CC in the given shift ranges (one such case is shown in  \autoref{correlation2} from \autoref{appendix:method}).  In \autoref{cor_coeff}, we show the cumulative distribution of the \ccmax\ for the Northern and Southern hemispheres in two latitude ranges (i) $\theta \leq 20^\circ$ and (ii) $\theta > 20^\circ$, depicting the fraction of \ccmax\ in various ranges. Furthermore, we also looked at the change in differential rotation parameter by changing the lower limit of \ccmax\ (see \autoref{appendix:method}), and we note that there is no significant change in rotation parameters after \ccmax=0.2. Hence, considering the potential negative impact of cases where \ccmax$<0.2$, they are discarded from our analysis. 

After discarding the cases mentioned above, we calculate the synodic rotation rate ($\Omega_{\rm synodic}$) using the $\Delta\phi$ corresponding to the \ccmax\ and $\Delta t$ as

\begin{equation}
    \Omega_{\rm synodic} = \frac{\Delta\phi}{\Delta t}.
    \label{eq6}
\end{equation}

To incorporate the effect of the motion of the Earth around the Sun, we apply a correction on the synodic rotation rate to get the sidereal rotation rate using the relation \citep{rosa1995, wittman1996, skokik2014}
\begin{equation}
    \Omega_{\mathrm{sidereal}} = \Omega_{\mathrm{synodic}} + \frac{\overline{\Omega}_{\mathrm{Earth}}}{r^{2}}\left(\frac{\cos^{2}\psi}{\cos i}\right),
    \label{eq7}
\end{equation}

where $\overline{\Omega}_{\mathrm{Earth}}$ is the mean orbital angular velocity of the Earth (0.9856 $^\circ {\rm /day}$), $i$ is the inclination of the solar equator to the ecliptic, $\psi$ is the angle between the pole of the ecliptic and the solar rotation axis orthographically projected on the solar disk, and $r$ is the Sun-Earth distance in Astronomical Units \citep[AU;][]{lamb2017, Jha2021}. Hereafter, we drop the subscript sidereal from $\Omega_{\mathrm{sidereal}}$ and use $\Omega$ instead for the same.

\section{Results}\label{Sec:results}

\subsection{Average Chromospheric Rotation Profile} \label{subsec:chromosphere}
\begin{figure*}[htb!]
 \centering
\includegraphics[width=\textwidth]{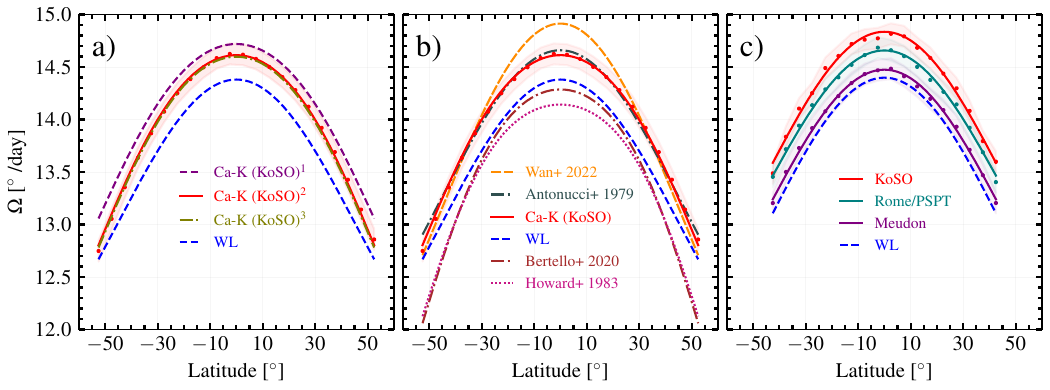}
\caption{a) The average rotation rate of the entire data period (1907\,--\,2007) calculated in each latitude band as a function of latitude along with the best-fit curve to the observed data represented as a solid red curve in the latitude range of $\pm55^{\circ}$. Dashed blue is the rotation profile of the photosphere using sunspot \citep{Jha2021}. The purple and olive curves are for the chromosphere rotation profile using KoSO data for two different periods: 1980\,--\,2007 and 1907\,--\,1979, respectively. b) Comparison between our results for KoSO data for the entire period and selected works from the literature.  c) A comparison between the resulting rotation profile of the chromosphere derived from different sources of \ca\ data over 2000\,--\,2002. In particular, we show results for KoSO (red), Rome/PSPT (teal), and Meudon (purple) \ca data. The dashed blue curve shows the rotation profile of the photosphere over the same period. \\
Note - $^{1}$ = (1980\,--\,2007) , $^{2}$ = (1907\,--\,2007), $^{3}$ = (1907\,--\,1979)
 }
\label{finalresult}
\end{figure*}

To get the latitudinal variation of the rotation profile, we calculate the mean of $\Omega$ for each latitude band. These mean $\Omega$ are shown by the filled red circles in \autoref{finalresult}(a) as a function of latitude ($\theta$). 
We performed two error estimations for our calculations: (i) the least count error ($\sigma_{\rm LCE}$) due to the resolution of the heliographic grid, i.e. $0.1/\Delta t$ ($\Delta \phi$  will have at least $0.1^{\circ}$ uncertainty) and (ii) the standard statistical error ($\sigma_{{\rm SSE}}$) of the mean. We calculate the combined errors in our analysis by the relation $\sigma_{{\rm total}} = \sqrt{(\sigma_{{\rm LCE}})^2 + (\sigma_{{\rm SSE}})^2}$. However, $\sigma_{{\rm LCE}}$ is dominant in the total error estimation, as $\sigma_{{\rm LCE}}$ is an order of magnitude greater than the $\sigma_{{\rm SSE}}$. Therefore, we find approximately the same errors (light red continuous band in \autoref{finalresult}(a) in all latitude bands. Now, we fit \autoref{eq1} to the mean $\Omega$ obtained using the Levenberg-Marquardt least square \citep[LMLS;][]{Marquardt2009} fitting method to get the differential rotation parameters $A$, $B$ and $C$ (see \autoref{tab:final}). The very first thing that we note is that our results suggest that the chromosphere plages give rotation rate $\approx1.59\%$ faster than the underlying photosphere, as inferred by using WL sunspot data \citep[blue dashed line in \autoref{finalresult}(a);][]{Jha2021}, which is in disagreement with the chromospheric rotation rate obtained in \citet[brown dashed-dotted in \autoref{finalresult}(b);][]{Bertello2020} using MWO \ca\ data. 
Interestingly, a very recent work \citep{li2023} reports a faster-rotating chromosphere using the auto-correlation technique on \ca\ synoptic maps from MWO. 
We will discuss more about the possible reasons behind this discrepancy as seen from MWO \ca\ plage data in \autoref{Sec:Discussion}.

We also observed a relatively higher rotation rate from the average value in the 2nd half period of KoSO data. As can be seen in  \autoref{finalresult}(a) and \autoref{tab:final}, the rotation rate averaged after 1980 (purple curve) is higher than the rotation rate averaged before 1980 (green curve) as well as the total average rotation profile (1907--2007; red curve). We suspect this may be due to the degraded data quality after 1980, as reported by \citet{Chatterjee2016, priyal2014, theo2019SoPh}.

There are two other works that attempt to measure the chromospheric rotation rate, \citet{antonucci1979} and \citet{wan2022SoPh}, using Ca network and \ca\ filaments, respectively, suggesting faster rotation of the chromosphere. These chromosphere rotation rates obtained in the past are overplotted in \autoref{finalresult}(b) to show a comparison with the rotation rate of the photosphere, measured by tracking sunspots \citep{Jha2021} and spectroscopic method \citep{howard1983}. Except \citet{Bertello2020} obtained from MWO data, all other results suggest that the chromospheric plages (and other chromospheric features) give a faster rotation rate than the photospheric rotation rate obtained from sunspot or surface rotation rate measured using spectroscopic methods.  

To further investigate these slightly different results obtained in the measured rotation rate, we implemented our image cross-correlation-based differential rotation measurement technique to the small sample of other data sets, which are discussed in the following section.

\begin{deluxetable*}{lllllll}
\tabletypesize{\scriptsize}
\tablewidth{0pt}
\tablecaption{ Solar differential rotation parameters from different observations \label{tab:final}}
 \tablehead{\colhead{Study} &\colhead{Data/Features} & \colhead{Observatory} &\colhead{Period} & \colhead{$A \pm\Delta A$ } & \colhead{$B \pm\Delta B$}& \colhead{$C \pm\Delta C$}\\
 \colhead{} & \colhead{}  & \colhead{}  & \colhead{} & \colhead{($^{\circ}\rm/day$)} & \colhead{($^{\circ}\rm/day$)} & \colhead{($^{\circ}\rm/day$)} 
 }
\startdata 
\citet{howard1983}    &Doppler Measurement&MWO      & 1967--1982  & 14.143 $\pm$ 0.006 & -1.718 $\pm$ 0.005 & -2.361 $\pm$ 0.007 \\
\citet{Jha2021}       &WL                 &KoSO     & 1923--2011 &  14.381 $\pm$ 0.004 & -2.72 $\pm$ 0.04 & --- \\
\citet{Bertello2020}  &\ca~plage          &MWO      & 1915--1985 &  14.2867 $\pm$ 0.0025 & -2.128 $\pm$ 0.0351 & -2.24 $\pm$ 0.0787 \\
\citet{wan_solar_2022}&\ca~plage          &MWO      & 1915--1985 &  13.496 $\pm$ 0.084   & -2.468 $\pm$0.656 &---\\
\citet{antonucci1979} &\ca~Network        &Anacapri & 1972 (8 May--14 August) &  14.66  & -2.79  & --- \\
\citet{wan2022SoPh}   &\ca~Filaments      &Coimbra  & 1929--1941 &  14.914 $\pm$ 0.263 & -3.505 $\pm$ 0.684 & --- \\
This work             &\ca~plage          &KoSO     & 1907--2007 & 14.61 $\pm$ 0.04 & -2.18 $\pm$ 0.37 & -1.10 $\pm$ 0.61 \\
This work             &\ca~plage          &KoSO     & 1907--1979 & 14.59 $\pm$ 0.04 & -2.23 $\pm$ 0.37 & -1.05 $\pm$ 0.60 \\
This work             &\ca~plage          &KoSO     & 1980--2007 & 14.72 $\pm$ 0.04 & -2.05 $\pm$ 0.39 & -0.94 $\pm$ 0.64 \\
\enddata
\tablecomments{Columns are the bibliographic entry, type of observation and feature used, name of observatory, period covered by data, and the parameters of fitting Eq. \ref{eq1}.}
%\label{tab:final}
\end{deluxetable*} 

\subsection{Comparison with Other Observatories} \label{subsec:otherobs}
To test the robustness of our algorithm and to ensure that the result that we are getting is not an artefact of the data, we implemented it on the \ca\ data obtained at Meudon \citep{malherbe2019} and Rome/PSPT \citep[][]{ermolli1998,ermolli10.3389} for the period of 2000\,--\,2002, which is close to the solar maximum and have significant plage regions. 
We applied the same process to determine the chromospheric rotation in the Meudon and Rome/PSPT data as we did for the KoSO ones.
In \autoref{finalresult}(c), we compare our results for the chromospheric rotation rate from KoSO, Meudon, and Rome/PSPT by using data only over the period 2000\,--\,2002. 
We find all three \ca datasets to result in differential rotations that are indeed faster than the one found for the photosphere by tracing sunspots \citep{Jha2021}. 
Additionally, it is crucial to acknowledge that, although the central wavelength of \ca\ filter for all these observatories is the same, they have different pass bands, e.g., KoSO: 0.05\,nm \citep{priyal2014}, Meudon: 0.015\,nm \citep[]{malherbe2019}, and Rome/PSPT: 0.25\,nm \citep{ermolli1998,ermolli10.3389}. 
Furthermore, KoSO and MWO data are spectroheliograms, while Rome/PSPT are filtergrams; thus, the shape of the pass bands might also differ.
Consequently, these data sets have contributions from slightly different layers of the chromosphere, and this might play a role in the
differences observed in the results. 
In \autoref{tab:other ca k}, we outline the differential rotation parameters from the best fit (\autoref{eq1}) for all these cases using the data over the period 2000\,--\,2002. 
However, we do not observe a monotonic change in differential rotation with archive bandwidth, thus height in the solar atmosphere.
Furthermore, in  \autoref{tab:other ca k}, we also note that the rotation rate obtained from the KoSO \ca\ data is relatively higher than the average rotation rate obtained from the same over the entire span of data (\autoref{tab:final}). The reason behind this observed higher value is already discussed in \autoref{subsec:chromosphere}.
We further extended the test of the robustness of our algorithm by implementing it on Michelson Doppler Imager \citep[MDI;][]{scherrer_solar_1995} intensity continuum data and MWO \ca\ data (discussed in \autoref{appendix:validation_whitelight}) to compare with the already obtained results in \citet{Jha2021} and \citet{Bertello2020}, respectively.

\begin{table}
\centering
\caption{Comparison between our results for the differential rotation from KoSO, Rome/PSPT, and Meudon \ca data to those from KoSO WL data by \citet{Jha2021} for the period 2000--2002.}
\begin{tabular}{rrrrrr}
\hline
\hline
    Data & $A \pm\Delta A$ & $B \pm\Delta B$ & $C \pm\Delta C$ \\
                & ($^{\circ}\rm/day$) & ($^{\circ}\rm/day$) & ($^{\circ}\rm/day$) \\
\hline
KoSO$^1$ & 14.39   & -2.83  & --- \\
KoSO$^2$ & 14.80 $\pm$ 0.05 & -2.13 $\pm$ 0.56 & -1.39 $\pm$ 1.25 \\
Rome/PSPT$^2$ & 14.64 $\pm$ 0.05 & -2.31 $\pm$ 0.57 & -0.68 $\pm$ 1.24 \\
Meudon$^2$ & 14.46 $\pm$ 0.04  & -2.39 $\pm$ 0.56 & -0.71 $\pm$ 1.22 \\
\hline
\end{tabular}
\tablecomments{\footnotesize{$^1$ \citet{Jha2021} for WL data; $^2$ This study \ca data}}
\label{tab:other ca k}
\end{table}

\subsection{North-South Asymmetry}

\begin{figure}[htbp!]
\centering
\includegraphics[width=\columnwidth]{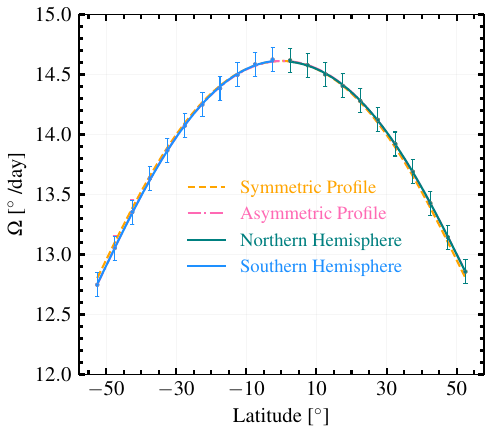}
\caption{The best-fit lines for the case of asymmetric fitting (\autoref{eq3}) in both hemispheres together (pink) as well as symmetric fitting (\autoref{eq1}) in both hemispheres together (yellow) and independently in Northern and Southern hemispheres (teal and dodgerblue, respectively).}
 \label{result_ns}
\end{figure}

There are various works \citep{schroter1975,Livingston1979,wan} that suggest that the chromosphere shows a significant difference in the observed rotation rate in the Northern and Southern hemispheres. Therefore, firstly, to investigate the North-South asymmetry in chromospheric rotation, we add the odd powers of $\sin{\theta}$ in \autoref{eq1}
\begin{equation}
\Omega= A + B^\prime\sin{\theta}+B\sin^2{\theta} +C^\prime\sin^3{\theta}+C \sin^4{\theta}.
\label{eq3}
\end{equation}
and again, fit this equation to the data using the LMLS method. Secondly, we use the \autoref{eq1} and fit it individually in the Northern and Southern hemispheres. In \autoref{result_ns}, we show all these three cases along with the fit of \autoref{eq1} in both hemispheres. When we fit the asymmetric profile (\autoref{eq3}), we note that the odd-order terms are very close to zero. Therefore, we can safely say that there is no or significantly less contribution from these terms in the chromospheric rotation profile, and it is symmetric within our precision of measurement. This is also confirmed by fitting \autoref{eq1} one by one in both hemispheres. The best fitting parameters are also summarized in \autoref{tab:asymmetry}.

\begin{figure}[htbp!]
\centering
\includegraphics[width=\columnwidth]{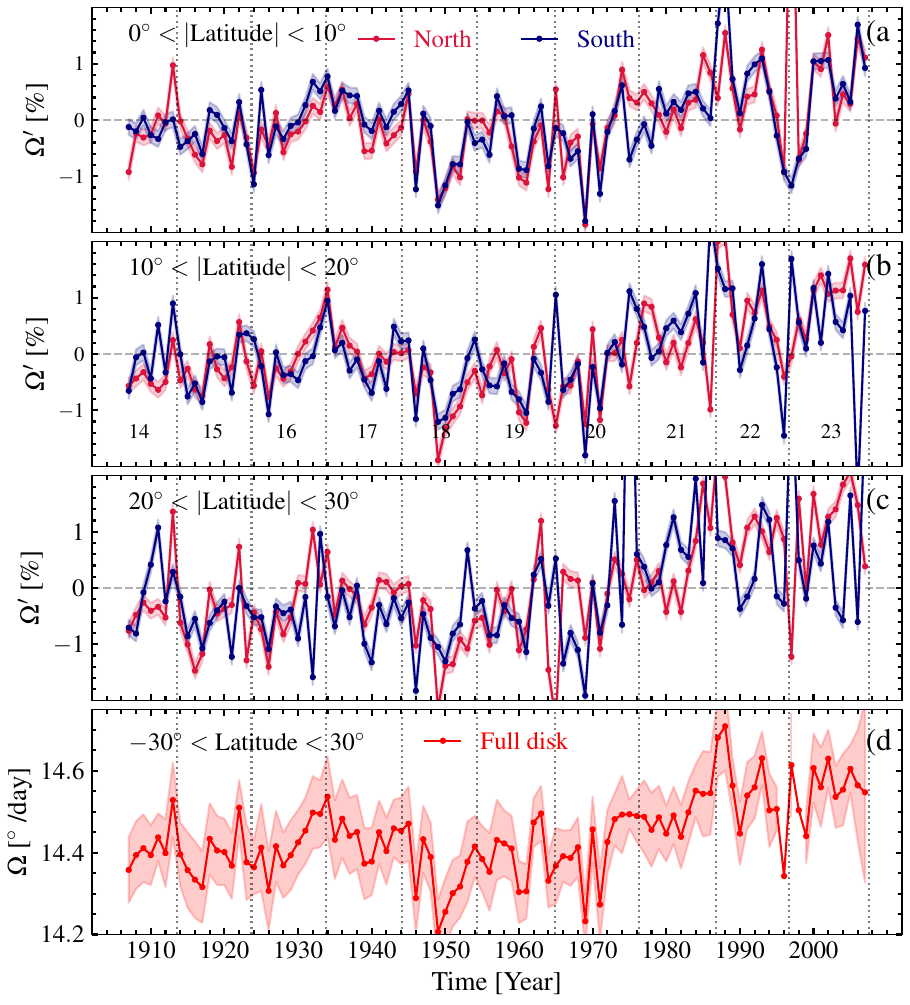}
\caption{Relative variations (\%) in rotation rate ($\Omega^\prime$) in three different latitude bands (a) $0^\circ$ \,--\, $10^\circ$, (b) $10^\circ$ \,--\, $20^\circ$ and (c) $20^\circ$ \,--\, $30^\circ$ for the Northern and Southern hemispheres and variation of $\Omega$ ($^{\circ}\rm/day$) for the years of 1907--2007 in panel (d).}
 \label{rel_change omega}
\end{figure} 

\begin{table*}
\centering
\caption{Comparison between the differential rotation parameters of Eq. \ref{eq3} for the Northern and Southern hemispheres.}
\begin{tabular}{rrrrrr}
\hline
\hline
        Results & $A \pm\Delta A$ & $B' \pm\Delta B'$ & $B \pm\Delta B$ &  $C' \pm\Delta C'$& $C \pm\Delta C$\\
                & ($^{\circ}\rm/day$) & ($^{\circ}\rm/day$) & ($^{\circ}\rm/day$) & ($^{\circ}\rm/day$) & ($^{\circ}\rm/day$)\\
\hline
Symmetric full disk& 14.61 $\pm$ 0.04 & --- & -2.18 $\pm$ 0.37 & --- & -1.10 $\pm$ 0.61 \\
Asymmetric full disk & 14.61 $\pm$ 0.04 &  0.02 $\pm$ 0.11 & -2.18 $\pm$ 0.38 & 0.07 $\pm$ 0.26 & -1.10 $\pm$ 0.61 \\
Symmetric Northern & 14.61 $\pm$ 0.06 & --- & -2.07 $\pm$ 0.53 & --- & -1.14 $\pm$ 0.86 \\
Symmetric Southern & 14.62 $\pm$ 0.06 & --- & -2.29 $\pm$ 0.53 & --- & -1.06 $\pm$ 0.85\\
\hline
\end{tabular}
\tablecomments{As symmetric fit we refer to Eq. \ref{eq1}, while as asymmetric to Eq. \ref{eq3}.}
\label{tab:asymmetry}
\end{table*}

We further investigate the difference in rotation rate in the Northern and Southern hemispheres by calculating the variation in rotation profile over the mean calculated from the entire duration of the data. Which we call a relative change in rotation rate and define as
\begin{equation}
\Omega^\prime=\frac{(\overline{\Omega}_{\mathrm{year}} - \overline{\Omega}_{\mathrm{all}})} {\overline{\Omega}_{\mathrm{all}}} \times 100\%,
    \label{relchangeequn}
\end{equation}
where $\overline{\Omega}_{\rm year}$ is the mean value over a year and $\overline{\Omega}_{\mathrm{all}}$ is the mean value over the entire period 1907--2007. In \autoref{rel_change omega}(a, b, c) we show the variation of $\Omega^\prime$ in three different bands (i) $0^\circ$\,--\,$10^\circ$, (ii) $10^\circ$\,--\,$20^\circ$, and (iii) $20^\circ$\,--\,$30^\circ$ for the Northern and the Southern hemispheres separately. We do note an upward trend in the rotation rate after 1980, consistent with the higher rotation rate obtained in \autoref{finalresult}(c), which is again the consequence of the degraded data quality in these periods \citep[see, e.g.][]{Ermolli2009ApJ,chatzistergos_analysis_2023}. Additionally, we also note this upward trend in \autoref{rel_change omega}(d) after 1980 in the yearly averaged $\Omega$, calculated by averaging over the disk (latitude range $-30^\circ$ to $+30^\circ$). Hence, the inference needs to be drawn very carefully.

\section{Discussion}\label{Sec:Discussion}

\begin{figure*}[htbp!]
\centering
\includegraphics[width=10cm]{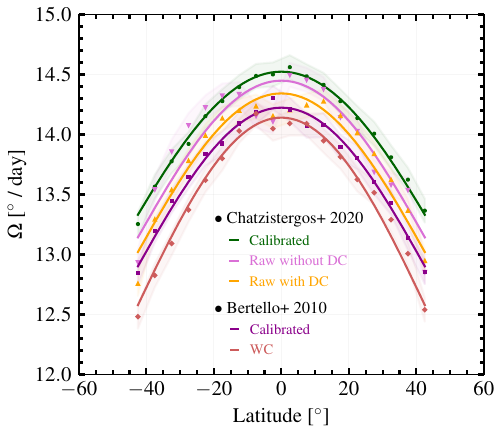}
\caption{Comparison of differential rotation profiles derived with different levels of processing of MWO \ca data over 1978--1979. 
The dark green and dark magenta curves show the rotation profiles using MWO data calibrated by \citet{chatzistergos_analysis_2020} and \citet{bertello2010SoPh}, respectively. 
The other three curves show raw scanned images without any processing to compensate for the limb darkening.
The yellow and pink curves are for the raw images used by \citet{chatzistergos_analysis_2020} after applying the Disk Circularisation (DC) and without it, respectively.
The red curve is for images  Without Calibration (WC) used by \citet{bertello2010SoPh}.
We note that the pink and red curves refer to data that differ in their spatial resolution (full size for pink, reduced size for red), have identified the solar disk differently, as well as differ in corrections that they applied in date/time and rotation angle.}
 \label{result_mwo_all}
\end{figure*}

One of the important aspects of the results obtained in this work is their dependence on various data sets, methods and calibration techniques. The results obtained based on the KoSO data are in contrast to \citet{Bertello2020} based on MWO \ca\ observations; at the same time, they are in agreement with the findings of \citet{antonucci1979,Li2020,wan2022SoPh}. We speculate that the difference in the results may be attributed to the differences in the applied processing of the images as well as the pass bands of the filters utilized at the respective observatories, leading to a dataset-specific dependence of the results. However, we must highlight that our understanding of such a dependence remains incomplete. Recently, \citet{li2023} utilized data from the MWO processed by \citet{Bertello2020} to arrive at a similar result to ours, thus contrasting that of \citet{Bertello2020}, which further underlines the complexity in the pursuit of a complete understanding of such a dependence. We extended our analysis by applying our method of differential rotation measurement on MWO data calibrated by two different techniques: one is done by \citet{bertello2010SoPh} and the other is done recently by \citet{chatzistergos_analysis_2020}. Here we note that we used full disk \ca observations processed by \citet{bertello2010SoPh} and not by \citet{Bertello2020}, on one hand because \citet{Bertello2020} provides only Carrington maps while \citet{bertello2010SoPh} gave the full disk images and on the other hand because the image processing applied in these two studies is effectively the same. Interestingly, we found a significant difference in our results (\autoref{mt.wilson and mdi}c, d) for these two data sets, which indicated that the different approaches of image processing might lead to different and even contradictory results. Differences in how the two processing approaches account for image distortions and the ellipticity of the recorded solar disk were found to contribute to this, but they are not the dominant factor (compare yellow and pink curves in Fig. \ref{result_mwo_all}). 
Residual artifacts in the images or issues with their orientation can also affect the efficacy of the process of estimating the rotation rate. 
In this direction, how the different calibration steps affect the results can also be seen in \autoref{result_mwo_all}. Here, we note that the curves listed as ``without DC'' (pink) and ``Without Calibration (WC)'' (red) are the datasets of raw images that were used by \citet{chatzistergos_analysis_2020} and \citet{bertello2010SoPh}, respectively. Both datasets include MWO images without compensation for the limb darkening or circularisation of the solar disk; however, they differ in their spatial resolution (pink has the full resolution images, while red is for the reduced size images used by \citealt{bertello2010SoPh}). 
Further differences exist because corrections in date/time information, as well as the orientation of the solar disk, were applied for the ``without DC" curve compared to the ``Without Calibration (WC)'' one.
This highlights the added uncertainty in the estimation of the differential rotation due to the spatial resolution and orientation of the images.
However, the exact reason behind the differences is still veiled to us, and this would require a more in-depth analysis.

\section{Summary and Conclusion}\label{Sec:summary}

We use full disk \ca\ (393.367\,nm) spectroheliograms from KoSO spanning a century (1907\,--\,2007; Cycles~14 to 23) to calculate the chromospheric differential rotation using a newly developed automatic algorithm. Our findings show that the average chromospheric rotation profile is
\begin{equation}
\Omega (\theta) = (14.61 - 2.18\sin^2{\theta} - 1.10\sin^4{\theta})^\circ{\rm /day}
\label{eq5}
\end{equation}
which is $\approx$1.59\% faster when compared with the photospheric equatorial rotation rate obtained by \citet{Jha2021} using WL KoSO data and $\approx$ 3.3\% compared to the results by \citet{howard1983} using MWO Doppler measurements (\autoref{finalresult}(b), \autoref{tab:final}).

Our analysis supplements previous results \citep[e.g.][]{Livingston1969,antonucci1979,Li2020,wan2022SoPh} suggesting that the higher layers of the solar atmosphere rotate faster than those underneath. We have validated our results by extending our method on the small samples of \ca\ data from other observatories such as Meudon, Rome/PSPT and MWO. Results obtained from these datasets extend affirmative support to our method as well as increase the reliability of our results. The contradiction of our result with that of \citet{Bertello2020}, as well as the contradiction between results obtained using two distinctly different methodologies \citep{bertello2010SoPh,chatzistergos_analysis_2020}, indicates a significant influence of image processing techniques as well as data-set specific factors. However, we must reiterate that the full extent of such an influence remains outside the scope of the current study.

In the past, based on these results, there have been attempts, such as that in \citet{weber1969}, to explain the observed increase in rotation rate with height based on the conservation of angular momentum in a magnetic field-dominated medium. However, it is important to acknowledge our constrained understanding of the matter, which further underlines a need for careful measurements of the rotation rate farther in the solar atmosphere at various heights above the photosphere. Therefore, in a follow-up study, we are attempting to achieve the same by measuring the solar rotation rate even higher in the atmosphere. We hope such a study will complement the current work and help us advance the broader understanding of the solar atmosphere.

\section{acknowledgments}
    We express our gratitude to the observers at the Kodaikanal Solar Observatory and the individuals involved in the digitization process for their efforts in providing extensive solar data spanning over a century to the scientific community. Kodaikanal Solar Observatory is a facility of the Indian Institute of Astrophysics, Bangalore, India. \ca\ raw data are now available for public use at \url{http://kso.iiap.res.in} through a service developed at IUCAA under the Data Driven Initiatives project funded by the National Knowledge Network. Additionally, we extend our sincere gratitude to the Meudon Solar Observatory \url{https://bass2000.obspm.fr/home.php}, Rome/PSPT \url{https://www.oa-roma.inaf.it/pspt-daily-images-archive/}, MWO \url{http://www.astro.ucla.edu/~ulrich/MW_SPADP} and SOHO/MDI \url{http://jsoc.stanford.edu/} for providing the easily accessible data that we have utilized in different segments of our current work. Meudon spectroheliograph data are courtesy of the solar and BASS2000 teams as part of the operational services of Paris Observatory. We acknowledge James and Mrinmoy Mukherjee for their support during the initial time of this work. The funding support for DKM's research is from the Council of Scientific \& Industrial Research (CSIR), India, under file no.09/0948(11923)/2022-EMR-I.
T.C. thanks ISSI for supporting the International Team 474 “What Determines The Dynamo Effectivity Of Solar Active Regions?”.
This research has made use of NASA's Astrophysics Data System (ADS; \url{https://ui.adsabs.harvard. edu/}) Bibliographic Services.

\bibliography{references}{}
\bibliographystyle{aasjournal}
\appendix

\section{Threshold in CC}\label{appendix:method}
\begin{figure}[ht!]
\centering
\includegraphics[width=12cm]{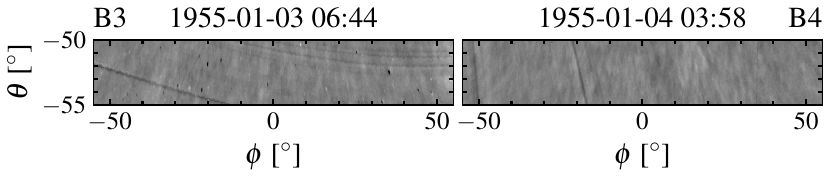}
\includegraphics[width=10cm]{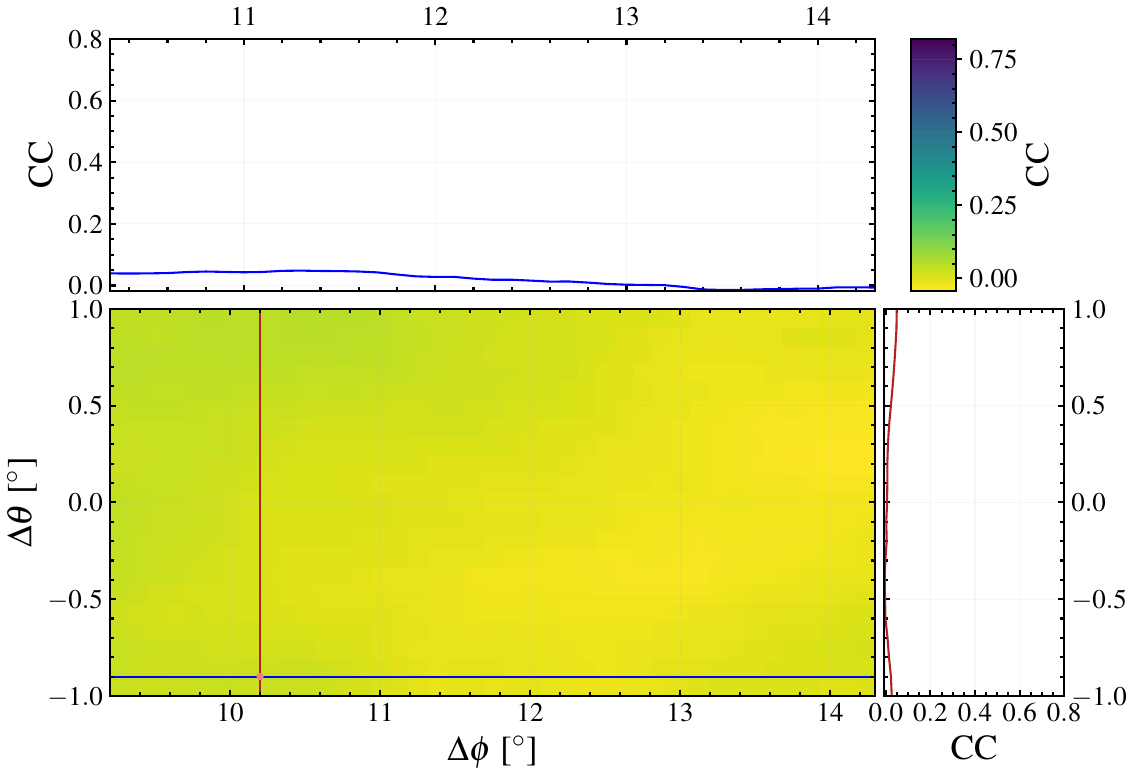}
%\caption{Same as the \autoref{correlation1}, but for different latitudinal range bins where plages are absent.}
\caption{The middle panel shows the 2D cross-correlation profile similar to \autoref{correlation1} but for latitudinal bins shown in B3 and B4 (in this case $-55^{\circ}$ to $-50^{\circ}$) where plages are absent.}
 \label{correlation2}
\end{figure}

\begin{figure}[ht!]
\centering
\includegraphics[width=10cm]{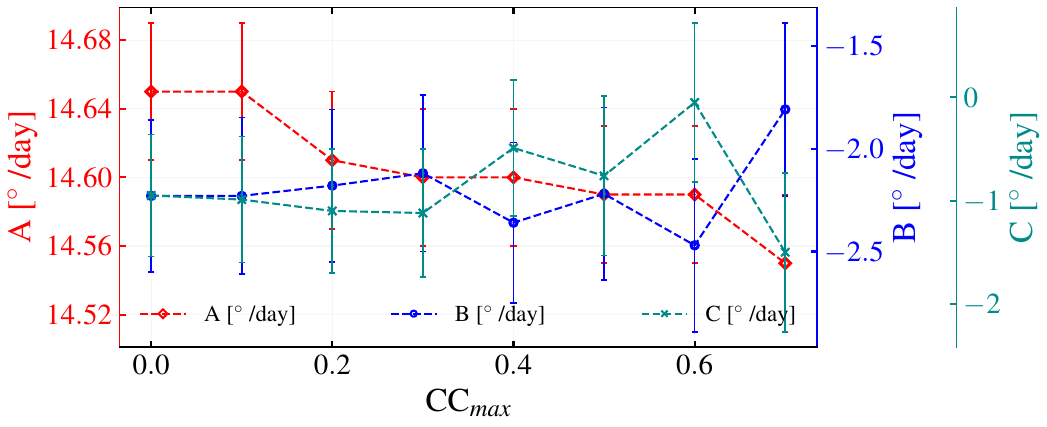}
\caption{The change in differential rotation parameters by varying \ccmax\ threshold.}
 \label{cor_coeff1}
\end{figure}

In this section, we explain the reason behind our choice of imposing a threshold in CC so as to not bias our results. Since, in our study, we have not pre-selected the best images or the latitude bands where plages are prominently present for our analysis, it may negatively impact our inference as our image correlation method is dominantly affected by the presence of plages and artifacts. In the latitudinal bands, where plages are mostly absent (see B3 and B4 in \autoref{correlation2}), we see either no/minute variation in CC in the cross-correlation matrix (one such example is shown in \autoref{correlation2}). There are cases where we see plages, still, we see similar behaviour of CC due to the poor image quality. Therefore, we only select the latitude range where we rarely expect the presence of plages (above 55$^\circ$ latitude). In addition to these, to calculate the best chosen $\Omega$ values, we have put a lower threshold on \ccmax\ of 0.2 in our analysis. In \autoref{cor_coeff1}, we notice that there is a sudden jump in differential rotation parameter (A) as we switch from 0.1 to 0.2, but no significant variation is seen after we increase the lower threshold from 0.2 to 0.7 in \ccmax\ (apart from the change in uncertainty because of poor statistics). Thus, to maximize the statistics and there is no significant change in the results after the \ccmax\ threshold value of 0.2, we decided to go with 0.2 as our lower limit on \ccmax. However, we emphasize that the reasons for low cross-correlation can be physical, such as the absence of plage regions mentioned above, but also technical, such as image distortions or artifacts unaccounted for by the processing techniques or even inaccurate orientation of the images. Restricting our analysis to locations where plage is found would completely miss these technical cases and have the potential to bias our results. The use of a threshold in the cross-correlation is a more robust way not to let such artifacts affect our results.

\section{Method Cross-Validation Using MDI continuum intensity and MWO Data }\label{appendix:validation_whitelight}

\begin{figure}[ht!]
\centering
\includegraphics[width=\columnwidth]{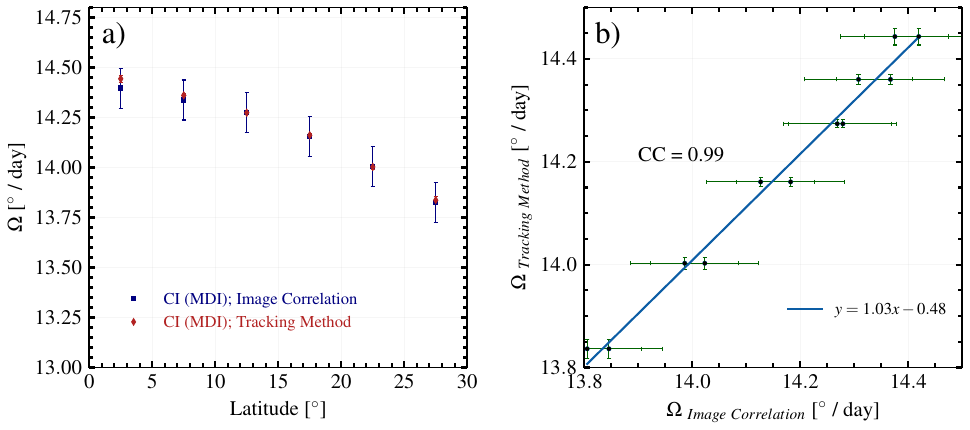}
\includegraphics[width=\columnwidth]{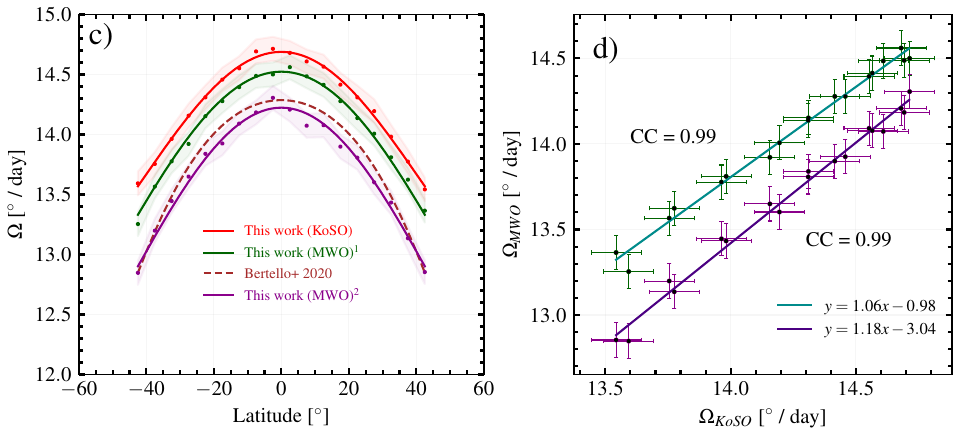}
\caption{(a) The photospheric rotation profile determined with the feature tracking method on MDI Continuum Intensity (CI) data and with the image cross-correlation technique, respectively. (b) The correlation plot between the angular rotation rate values obtained from the tracking method and the image cross-correlation technique. (c) The rotational profile of the chromosphere using MWO data for the period (1978\,--\,1979). The red curve shows the rotational profile for the KoSO \ca\ data (1978\,--\,1979), while the green curve and magenta represent the MWO rotation profile using data calibrated by \citet{chatzistergos_analysis_2020} and by \citet{bertello2010SoPh}. The brown curve represents the results obtained in the past done by \citet{Bertello2020}. (d) A correlation plot between the angular rotation rate values obtained from KoSO data and the rotation rate from MWO data.\\
Note - \citet{chatzistergos_analysis_2020}$^{1}$, \citet{bertello2010SoPh}$^{2}$}
 \label{mt.wilson and mdi}
\end{figure}

The Michelson Doppler Imager \citep[MDI;][]{scherrer_solar_1995} instrument on the Solar and Heliospheric Observatory (SOHO) satellite has provided a comprehensive dataset of Continuum Intensity (CI) observations spanning 15 years (1996\,--\,2011)\footnote{Data is available at \url{http://jsoc.stanford.edu/}}. In a recent study conducted by \citet{Jha2021}, the rotation profile of the solar photosphere was determined through the tracking of sunspots by utilizing MDI CI data. Taking advantage of the pre-established results derived from MDI CI data for the photosphere's rotation profile, we applied our image correlation methodology to verify its consistency. Our analysis, as depicted in \autoref{mt.wilson and mdi}(a), illustrates an overlap between the rotation profiles obtained via the two approaches (blue square; image correlation and red diamond; sunspot tracking). Furthermore, the scatter plot of angular velocity ($\Omega$) values obtained through these two methods, as presented in (\autoref{mt.wilson and mdi}(b)), demonstrates a Pearson correlation coefficient of 0.99. This high correlation coefficient serves as compelling evidence substantiating the validity and robustness of our methodology.

Mount Wilson Observatory (MWO) has observed \ca\ images (spectroheliograms; 0.035\,nm pass band filter) of the Sun from 1915 to 1985. The study of chromospheric differential rotation was done on MWO \ca\ data (1915\,--\,1985) by \citet{Bertello2020} and it was found that the chromosphere rotates slower compared to the photosphere (discussed in \autoref{Sec:results}). To verify our method, we followed the same procedure as our method for the newly calibrated data done by \citet{theo2018,theo2019,chatzistergos_analysis_2020} and for the data calibrated by \citet{bertello2010SoPh}\footnote{the data are from \url{http://www.astro.ucla.edu/~ulrich/MW_SPADP}}. For the data of \citet{bertello2010SoPh}, we used the resolution of the heliographic grid as 0.25 instead of 0.1 because the mentioned data have spatial pixels of 866\,pixels\,$\times$\,866\,pixels. Here, we must note that the data calibrated in \citet{bertello2010SoPh} are further used in \citet{Bertello2020} for the measurement of the MWO rotation rate. 
Surprisingly, we observed a significant difference in the results obtained from both data, as can be seen in \autoref{mt.wilson and mdi}(c). 
The rotation rate acquired from MWO data processed by \citet[green curve;][]{chatzistergos_analysis_2020} is higher than the result by \citet[brown curve;][]{Bertello2020}, and closer to our result for KoSO data (red curve). 
However, our estimate of differential rotation with MWO analyzed by \citet[magenta curve;][]{bertello2010SoPh} suggests a slightly slower rotation rate than the result by \citet[brown curve;][]{Bertello2020} for the same data. 
Various things contribute to these differences, including the different approaches in preprocessing and calibrating the MWO data, as well as in the process of determining the differential rotation. 
But what particular step in the preprocessing and calibration is affecting is still not understood. 
Also, we plotted the scatter plot of both $\Omega$ values and got the Pearson correlation coefficient of 0.99 \citep{chatzistergos_analysis_2020} and 0.98 \citep{bertello2010SoPh} that can be seen in (\autoref{mt.wilson and mdi}(d)) which shows the strong correlation between KoSO and MWO data set.

\end{document}